\documentclass[onecolumn,showpacs,preprint2,amsmath,amssymb]{revtex4}

\usepackage{epsfig}
\usepackage{bm}

\begin{document}

\title{Neutrino-Nucleus Cross Section Measurements using Stopped Pions 
and Low Energy Beta Beams}
\author{G. C. McLaughlin}
\email{Gail_McLaughlin@ncsu.edu} 
\affiliation{Department of Physics, North Carolina State University, 
Raleigh, North Carolina 27695-8202}

\begin{abstract}

Two new facilities have recently been proposed to measure low energy
neutrino-nucleus cross sections, the $\nu$-SNS (Spallation Neutron Source) 
and low energy beta beams. The
former produces neutrinos by pion decay at rest, while the latter
produces neutrinos from the beta decays of accelerated ions.  One of the
uses of neutrino-nucleus cross section
 measurements is for supernova studies, where typical neutrino
energies are 10s of MeV.  In this energy range there are many different
components to the nuclear response and this makes the theoretical 
interpretation of the results
of such an experiment complex. Although even one measurement on a heavy nucleus
such as lead is much anticipated, more than one data set would be 
still better.
We suggest that this 
can be done by breaking the electron spectrum down into the parts
produced in coincidence with one or two neutrons, running a beta beam at
more than one energy, comparing the spectra produced with pions and
a beta beam or any combination of these.

\end{abstract}

\maketitle

\section{introduction}

Neutrino-nucleus cross section measurements are desirable 
from the point of view of understanding nuclear structure, but they
are perhaps even more desirable for astrophysical 
reasons.  The supernova is the best studied 
astrophysical environment where neutrino
scattering reactions have significant impact.
Proper inclusion of the reverse process, 
electron capture on nuclei, has recently been shown to factor significantly
in the prospects for obtaining a supernova explosion \cite{Hix:2003fg}.  
Furthermore, neutrino-nucleus
interactions figure heavily in determining the nucleosynthesis that
is produced during the course of a supernova explosion.  
Neutrino nucleosynthesis,
which occurs when neutrinos spall neutrons and protons off of 
pre-existing nuclei, is driven entirely
by neutrino-nucleus interactions 
\cite{Woosley:1989bd,Heger:2003mm}.  Several papers have
suggested that the r-process of nucleosynthesis
may be impacted heavily by neutrino-nucleus interactions,
e.g. \cite{McLaughlin:1996eq,Haxton:1996ms}. In fact 
these 
reactions may have such a detrimental effect that they are an effective
tool in constraining the environment \cite{Meyer:1998sn}.  
Thirdly,  
neutrino-nucleus measurements are needed to calibrate supernova neutrino
detectors.  For
a description of supernova neutrino detection using lead,
see \cite{Boyd:em}.  A recent review of different techniques
used to calculate neutrino-nucleus cross section measurements 
is given in \cite{Kolbe:2003ys}.

Traditional neutrino beams are created using pions which produce
both muon neutrinos and muon antineutrinos, and either electron
neutrinos or electron antineutrinos. This is the case for the
proposed $\nu$-SNS which will produce neutrinos from pions
decaying at rest by way of the Spallation Neutron
Source at Oakridge National Laboratory \cite{nuSNS}.  This facility will
make improvements on existing measurements of nuclei such as carbon and iron
\cite{Krakauer:1991rf,Armbruster:vd,Berge:1989hr} and measure cross 
sections on new nuclei such as lead.

Newly proposed beta beam facilities \cite{Zucchelli:sa}
produce either electron neutrinos or antineutrinos from beta-plus
or beta-minus decays of radioactive ions.   
Feasibility studies for beta beams 
are underway and a design is discussed in 
\cite{Lindroos:2003kp,Bouchez:2003fy}.
Beta beams were originally proposed 
as a way to make high energy neutrinos for use in
 long-baseline studies to determine the third, and as yet unknown
mixing angle in the neutrino mixing matrix,
 $\theta_{13}$ and to investigate CP-violation in the lepton sector
\cite{Mezzetto:2003ub,Bouchez:2003fy}.  However, lower energy beta beams have 
been proposed by Volpe \cite{Volpe:2003fi} and an application for neutrino
magnetic moment measurements has been discussed in \cite{McLaughlin:2003yg}.

In this paper we consider neutrino-lead measurements using spectra 
that would be produced by the $\nu$-SNS and 
from a beta beam facility. 
We consider a target mass of
10 tons, 20 meters away from the pion source and 10 meters away from the end
of a straight section of a beta beam ring.
We examine the spectra of the electrons that would be produced
from charged current interactions.

We explore the signals which can be produced if 
the electrons can be identified as being created in coincidence with 
zero, one or two spalled neutrons. We also suggest the possibility that
low energy beta beams be operated at more than one energy, therefore
producing different neutrino energy spectra.

\section{Neutrino Spectra}

In order to explore the effect of using 
different neutrino spectra to probe the
 nuclear response of $^{208}{\rm Pb}$, we must first compare the various
neutrino spectra themselves.  We are particularly 
interested in the charged current interactions of
electron neutrinos, since lead is largely Pauli blocked
in the electron antineutrino capture direction, and muon neutrinos
and antineutrinos are not energetic enough to produce muons for the
energy range we consider here.  

Electron neutrinos coming from stopped pions are produced by the 
decay of pions into muons which then decay into electrons,
$\pi^+ \rightarrow \mu^+ \nu_\mu$, $\mu^+ \rightarrow e^+ \nu_e \bar{\nu}_\mu$.
Electron neutrinos produced with  a low energy beta beam would come from 
beta-plus decay of a radioactive nucleus such as $^{18}{\rm Ne} \rightarrow
^{18}{\rm F} + e^+ + \nu_e$. Neon-18 decays in its rest frame 
with a half life $\tau_{1/2} = 1.67$ seconds 
and has a difference of nuclear masses of $Q_n = 3.93$ MeV, and therefore
the maximum energy for the neutrinos is about 3.4 MeV.  This 
energy is lower than that of most neutrinos that are produced in a supernova. 
These have approximately thermal spectra and average energies between 10 and
25 MeV although the exact values depend on the model, and the subsequent 
neutrino mixing.
However if Neon-18 were boosted to $\gamma \sim 10$ or even $\gamma \sim 5$ 
then the spectrum
falls in the same range as that of supernova neutrinos.

\begin{figure}[t]
\vspace*{-2cm}
\begin{minipage}{8cm}
\includegraphics[angle=0,width=6cm]{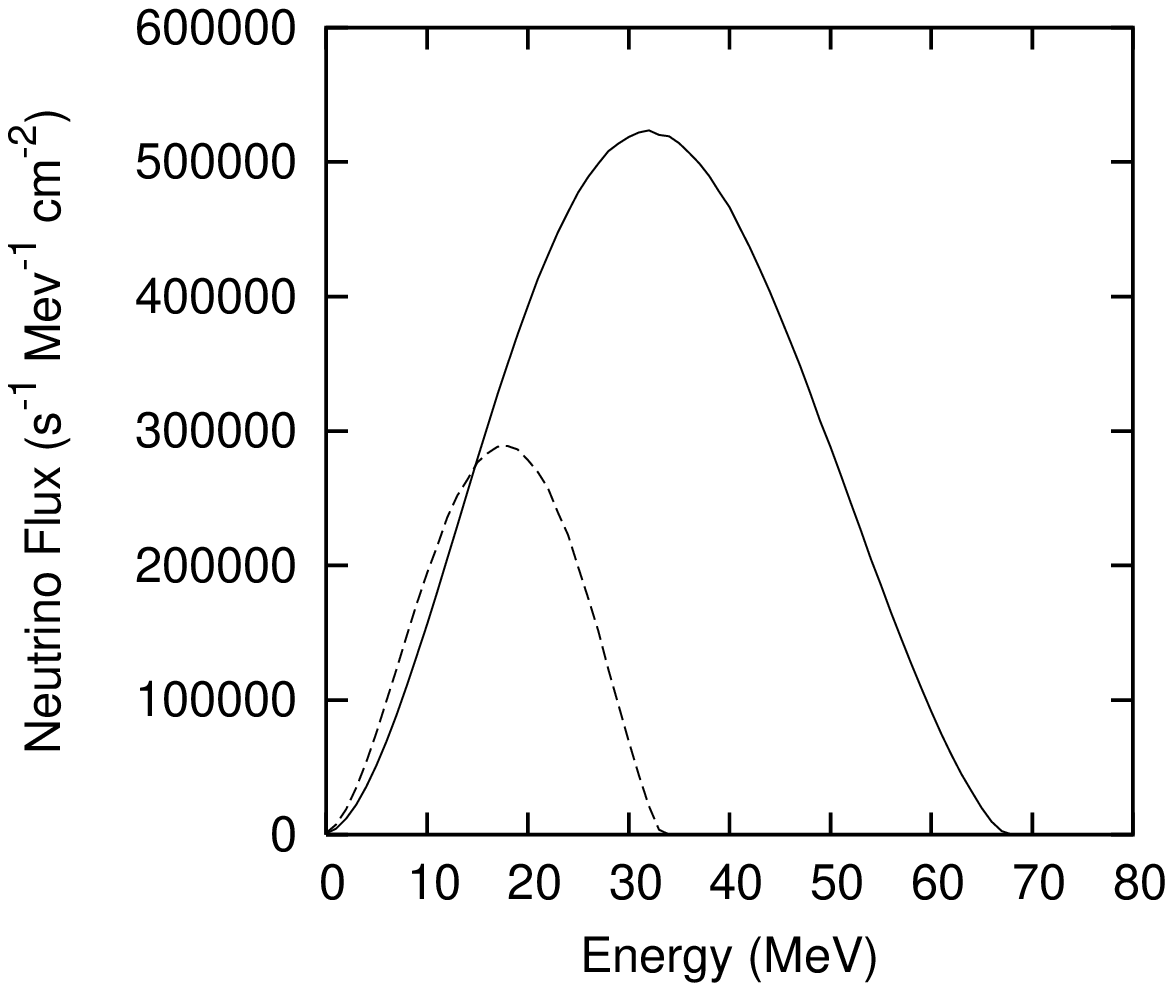}
\caption{{\sc Beta Beam, $\gamma=5$ and $\gamma=10$:} 
Figure shows the boosted spectrum of $\nu_e$s 
produced at a rate of $10^{13} s^{-1}$ 
from $^{18}{\rm Ne}$.  Calculated for
a target that has a cross sectional area of 4m
and is located 10m from one end of the ring, which has a straight
side length of more than 90m.  These numbers are further reduced by the
fraction of time the ions spend on the straight section of the ring.
\label{fig:betaboost}}
\end{minipage}\hspace*{0.5cm}
\begin{minipage}{8cm}
\includegraphics[angle=0,width=6cm]{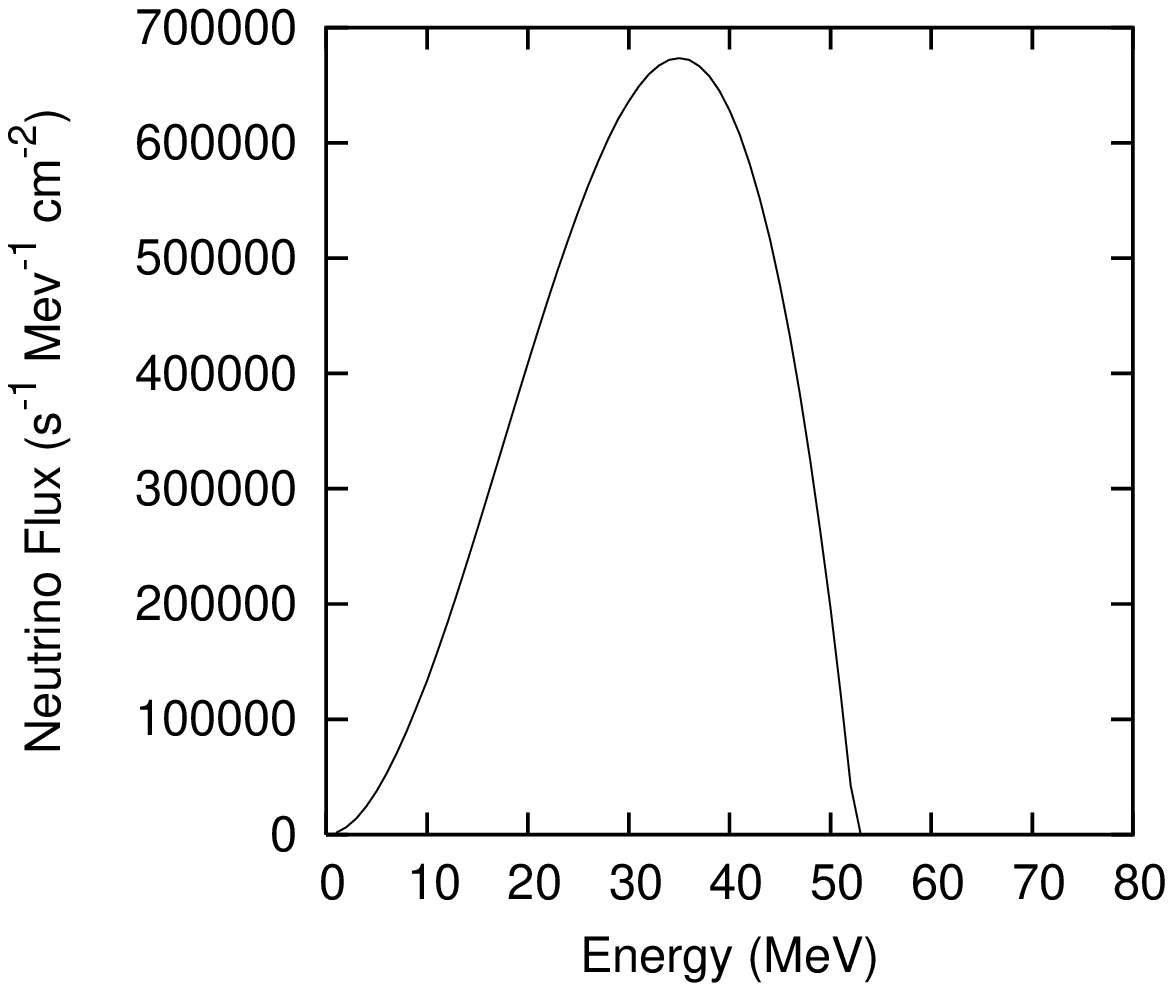}
\hspace*{0.5cm}
\caption{{\sc Electron neutrinos from stopped pions}:   
Figure shows the spectrum of $\nu_e$s from a $10^{15} {\rm s^{-1}}$ neutrino
source coming from stopped pions.  The target is assumed to be 20 m away
with a cross sectional area of 4 m$^2$  \label{fig:pionnuspect}}
\end{minipage}
\end{figure}

In Figs. \ref{fig:betaboost} and \ref{fig:pionnuspect} we show three different 
electron neutrino spectra.  One comes
from stopped pions, another from a Neon-18 beta beam boosted to 
$\gamma = 10$ and a third with a lower boost 
of $\gamma = 5$.  Although for a particular
target geometry and set up spectra will need to be calculated more precisely,
we can obtain a general idea of the relative effects of these different
sources using the following rough estimates. In the case of the pions  
we assume that the ten ton target 
has a cross sectional area of 4 m$^2$ and is located
20 meters from the source of the neutrinos.  For the beta beams we assume
that the same target is located 10 meters from the end of a straight section
in the ring of 90m in length.
The spectrum for the beta beams takes on a different shape and magnitude
depending on how large an opening
angle the detector subtends. The opening angle for the beam is
the about 7 degrees right as the ions reach 10 meters from the
detector, although the opening angle, and therefore the flux falls
off quickly as the distance from the source increases. 
  
For the pions we take
 a decay rate such that $10^{15} \, {\rm s}^{-1}$ electron
neutrinos are produced, which is the order of magnitude 
 discussed in the $\nu$-SNS proposal \cite{nuSNS}.
For isotropically emitted neutrinos, the
number hitting the target is reduced by 
$\sim 4 {\rm m}^2 /[ 4 \pi (20 {\rm m})^2]  = 0.08\%$ for this
geometry. 
For the beta beams we take a production rate of $10^{13}$ ions per second 
as discussed in 
\cite{Lindroos:2003kp}.  However, the
beam is collimated, since the relativistic boost will shift the
neutrinos toward the forward direction. Some estimates can be made
by starting with isotropically emitted neutrinos in the rest frame
of the ion,  boosting the component of neutrino 
momentum along the beam line, calculating the lab frame angle 
for the neutrino and determining if it is less than 
$\theta = \tan^{-1}(R/L)$, where R is the radius of the  circular
cross section of the detector and L is the distance of the neutrino
from the target.  For $\gamma=10$, the flux entering
the detector falls to 1.5\% when the ions are 100 meters from the target
and the integrated fraction of the flux which enters the
target that has been emitted 
between 10 meters and 100 meters is 8\%.  Furthermore,
for $\gamma=5$, the flux entering the detector falls to 0.3\% when
the ions are 100 meters away from the detector and the integrated 
fraction emitted from
between 10m and 100m is 2\%.  Even for a larger boost
of $\gamma = 15$, the fraction entering the detector is only 3.5\% at 100m
away, and the integrated fraction is 14\%.  Longer
straight sections therefore do not help much to increase the flux.  Since
all the numbers must be further reduced by the fraction of the entire
ring of which these 90 meters consist, 
smaller rings are advantageous.  A more complete discussion
of how ring and detector geometry influences fluxes and total event rates
is given in \cite{newvolpe}.

It can be seen from the figures that although the spectra are of the
same order of magnitude in 
energy, they do not have the same shape or even the same average energy
 and will therefore produce a different signal from
the same target nucleus.

\section{Electron Spectra}

The next step is to investigate the electron signal produced from the 
neutrinos interacting with the lead target.  In particular we 
investigate $\nu_e + ^{208}{\rm Pb} \rightarrow ^{208}{\rm Bi} + e^-$.
The Bismuth may produce one or more spalled neutrons as a result of
this charged current reaction.  Some neutrons will also be produced
from neutral current interactions as well, but they are not investigated
here. Natural lead
is a primarily $^{208}{\rm Pb}$ but it contains also a significant
amount (almost half) of other isotopes.  In this paper we use
pre-existing calculations of $^{208}{\rm Pb}$, as no calculations
of other isotopes exist to date. Although they are expected
to be similar, a complete anaylsis of a neutrino-lead experiment
would involve all nuclei present in the target. 
There are many papers which calculate the 
cross section of $^{208}{\rm Pb}$
\cite{Jachowicz:2003iz,Fuller:1998kb,Kolbe:2000np,Volpe:2001gy,Engel:2002hg}. 
We use cross sections from 
\cite{Engel:2002hg} in order to illustrate what
can be expected from such experiments.
All calculations agree at low energy, although at high energy
they begin to diverge.  This is because the nuclear response at low
energy transfer primarily consists of 
allowed transitions which are best understood. However, even for the allowed
transitions, there are open questions, such as 
the relative weight of the vector and axial
vector couplings in the nucleus.
The results of any neutrino-nucleus cross section experiment will be
used to calibrate theory and differentiate between different calculations.  

\begin{figure}[t]
\vspace*{-2cm}
\begin{minipage}{8cm}
\includegraphics[angle=0,width=6cm]{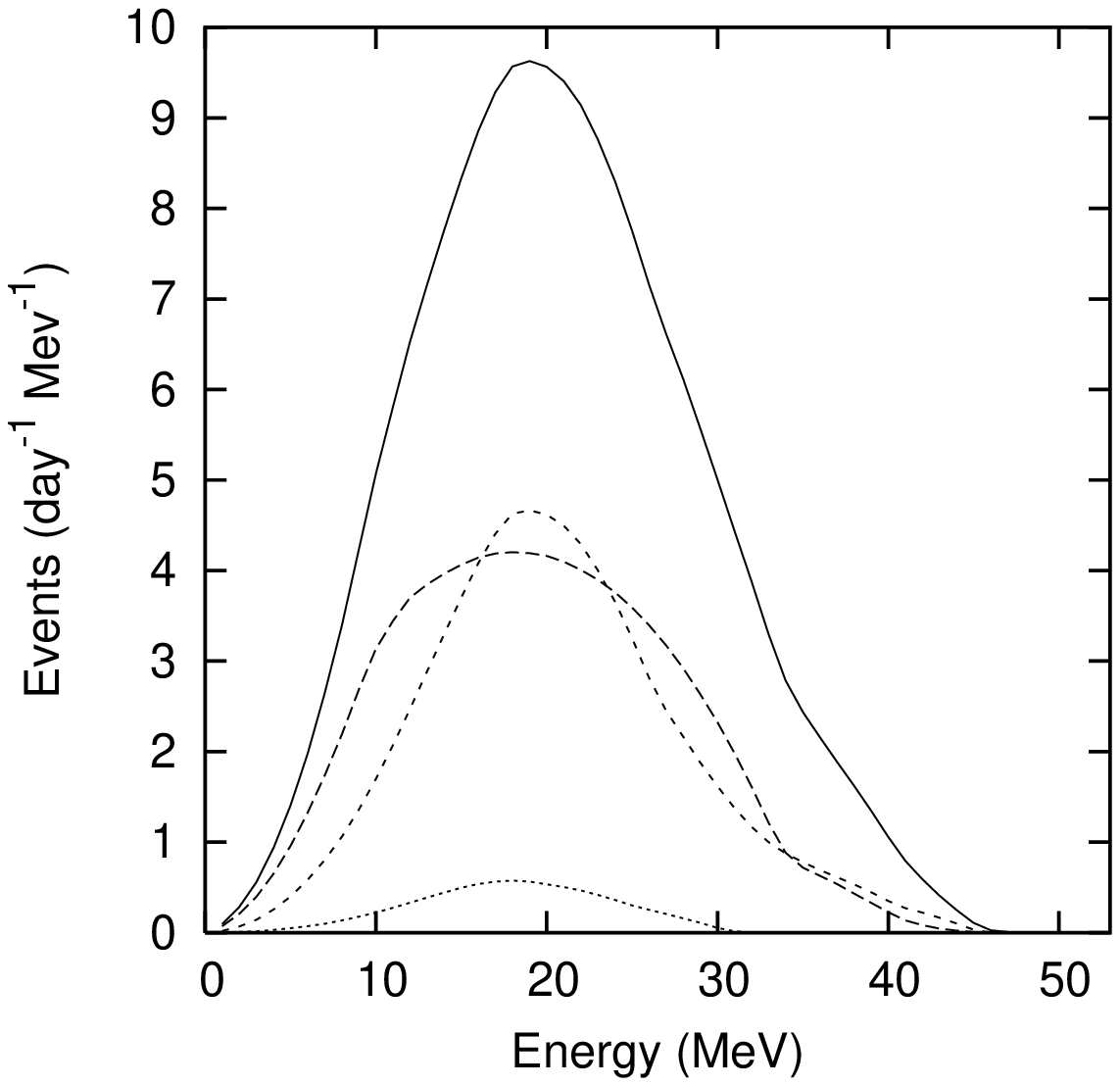}
\caption{{\sc Electron Spectrum, Pion Source:} 
Shows the total electron spectrum (solid line) coming from
electron neutrino capture on lead. Also shows the allowed
contribution (long dashed line), the $0^-$,$1^-$ and $2^-$ contribution
(short dashed line) and the $2^+$,$3^+$,$3^-$,$4^+$ and $4^-$ contribution
(dotted line).
\label{fig:crosspiondecomp}}
\end{minipage}\hspace*{0.5cm}
\begin{minipage}{8cm}
\includegraphics[angle=0,width=6cm]{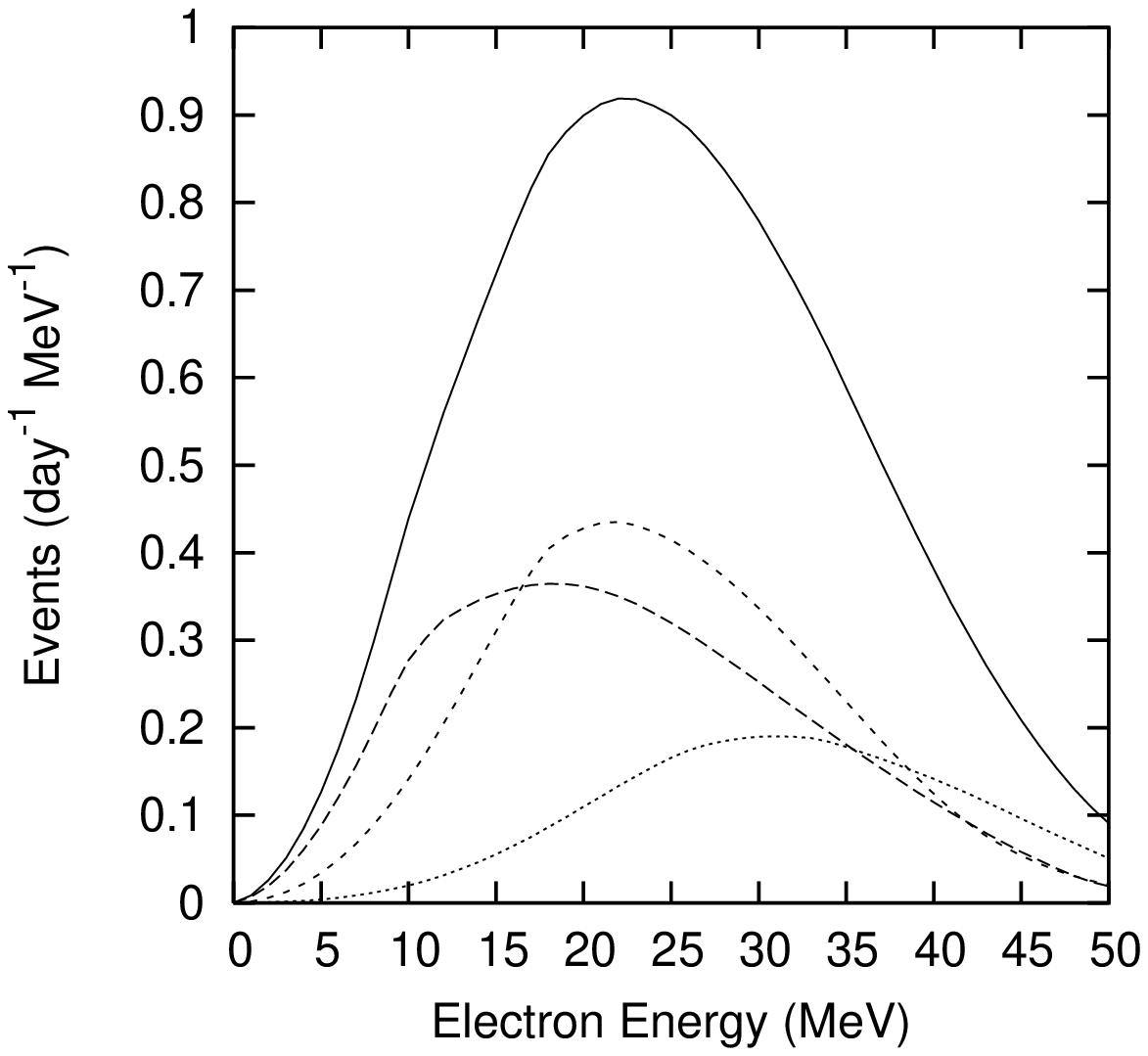}
\hspace*{0.5cm}
\caption{{\sc Electron Spectrum, $\gamma=10$}:   
Same as Fig. \ref{fig:cross10decomp} except for beta
beam ions accelerated to $\gamma =10$.  Different parts
of the nuclear response contribute at different
levels than for the $\nu_e$s from pions. \label{fig:cross10decomp}}
\end{minipage}
\end{figure}

We first plot numbers of electron events as a function of energy for
the stopped pion source.  This is shown in Fig \ref{fig:crosspiondecomp}.
In Fig. \ref{fig:cross10decomp}  we show the number of electron events for 
the $^{18}{\rm Ne}$, $\gamma = 10$ beta beam.  
As can be see from Table
\ref{tab}, the total number of 
electron events, using 100\% efficiency in each case is 
200 per day for the pion source and 
26 for the beta beam at $\gamma=10$. For this table, the numbers of
electron 
events for the beta beam can be roughly scaled as
\begin{equation}
N = \left({N_\nu \over 10^{13} {\rm s}^{-1}}\right) 
\left( {900 {\rm m} \over L}\right) 
\left({M \over 10 \, {\rm tons}}\right) N_{\rm table},
\end{equation}
where $M$ is the mass of the target, $L$ is the total length of
the ring and $N_\nu$ is the number of neutrinos emitted per second in
the ring. However,
changing the cross sectional area of the detector and the distance of the
detector from the beam is a nonlinear effect since it alters the spectra
through the angular dependence of the Lorentz boost. 

In both of Figs \ref{fig:crosspiondecomp} and \ref{fig:cross10decomp}
  we
show the contribution from various parts of the nuclear response, allowed
($0^+$ and $1^+$) and different parts of the forbidden.  As can be seen, 
different parts contribute with different
weight at different energies to the signal. Understanding which pieces
contribute with which weight is an important aspect of the theoretical
interpretation of the signal.  It is
 necessary in order to be able to accurately
calculate supernova neutrino-nucleus scattering where the spectra are
different still and will help to understand such theoretical questions
as which underlying force fits the response best.

Although any electron spectrum produced from neutrino-lead scattering
would be a great improvement on the
current experimental situation, to understand which nuclear transitions
are producing given electrons would be better still.
Although there
is no way to determine experimentally  whether a particular electron 
was produced by way of a particular  nuclear matrix element, 
there are several ways to obtain more information in this direction.  
One way is to
separate out the electrons that are emitted in
coincidence with zero, one or two neutrons. Another way would be
to compare a stopped
pion signal to a beta beam signal and fit both simultaneously.  
Finally, one could run a beta beam at
different energies, boosting for example to $\gamma = 10$ and then
 to $\gamma =15$ or $\gamma =5$.

\begin{figure}[t]
\vspace*{-2cm}
\begin{minipage}{8cm}
\includegraphics[angle=0,width=6cm]{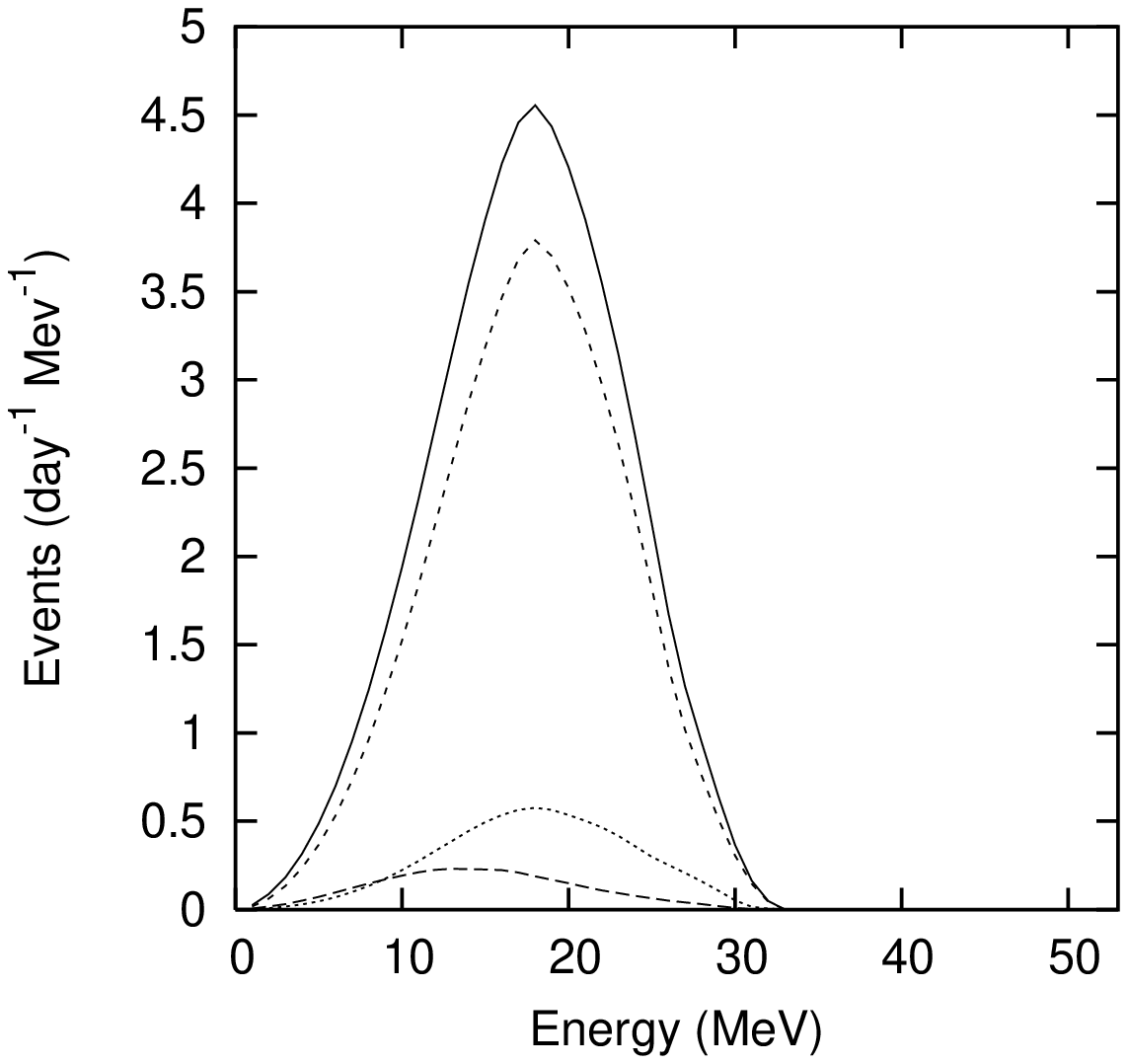}
\caption{{\sc Electron Spectrum, Pion Source:} 
The lines are the same as in
 Fig. \ref{fig:crosspiondecomp}, although only electrons
associated with two neutrons are shown. Most of the
electrons are produced from the forbidden 0-,1- and 2- transitions.
 \label{fig:crosspion2ndecomp}}
\end{minipage}\hspace*{0.5cm}
\begin{minipage}{8cm}
\includegraphics[angle=0,width=6cm]{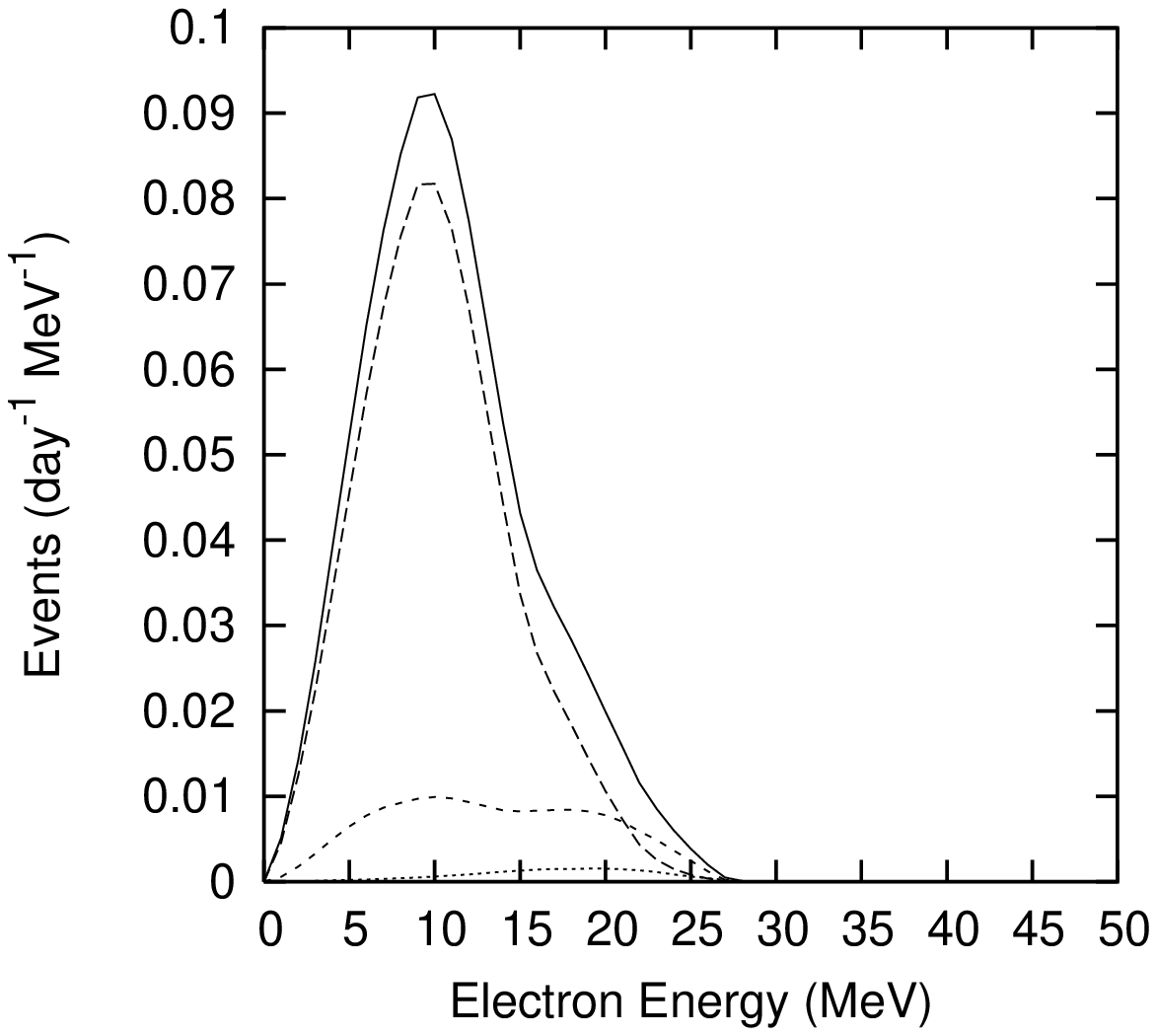}
\hspace*{0.5cm}
\caption{{\sc Electron Spectrum, $\gamma=5$}:   
Same as Fig. \ref{fig:cross10decomp} except for ions
accelerated to smaller energy. Most of the electrons
are produced from allowed nuclear transitions.  
 \label{fig:cross5decomp}}
\end{minipage}
\end{figure}

In Fig. \ref{fig:crosspion2ndecomp}  we show the electron energy spectrum for
only those electrons emitted with two neutrons, for the pion decay
at rest source, although the same can be done for the beta beam. 
There are 66 events per day associated with two neutrons for the pion source
and 10 events per day 
associated with two neutrons for the beta beam with $\gamma = 10$.
In Fig. \ref{fig:crosspion2ndecomp}, the  majority
of the electrons, 80\%,  are produced by the $0^-$,$1^-$ and $2^-$ 
part of the signal. 
In a similar way, the electrons associated
with one neutron can be shown to come primarily, 75\%, from the allowed part
of the cross section.
As can be seen
from a comparison with Fig. \ref{fig:crosspiondecomp}, much more
information can be obtained if the electrons can be identified as being 
produced with either one or two neutrons, than from the aggregate spectrum 
alone.  Such a two neutron analysis
would also be particularly useful when applied to a lead
based supernova neutrino detector \cite{Fuller:1998kb}
such as OMNIS or LAND 
\cite{omnis,land}, since
supernovae electron
neutrinos will oscillate with the muon and tau type neutrinos, and
the two neutron signal will consist primarily of electron neutrinos
that were originally prouced as muon neutrinos. If electron neutrinos
and muon neutrinos from the supernova are widely separated in energy, then 
observing the two neutron signal would be an effective
way to ``measure'' the original muon or tau spectrum \cite{Engel:2002hg}.

In Fig. \ref{fig:cross5decomp} 
we show the electron spectrum associated with a beta beam
boosted to $\gamma=5$ instead of $\gamma=10$.  For the lower energy
spectrum there
are 1.1 total electron events per day, and only 0.05 
electrons associated with two
neutrons.  The
majority of electrons 
are produced by way of allowed transitions in the nucleus.
Some information about these allowed transitions can be inferred
from (p,n) reaction data \cite{pn}.  However, a beta beam of this energy
 would provide a first
nearly direct measurement of these transitions, without the need to
decompose the signal by coincident number of neutrons.  
It can also provide
information about the relative strength of the $0^+$ or Fermi transition 
(governed by $g_V$) and the $1^+$ or Gamow-Teller transitions 
(governed by $g_A$).  
If it were possible to
take data at a boost factor of both $\gamma=5$ and $\gamma=10$, then
the information obtained by the $\gamma=5$ measurement could be used to 
calculate the allowed part (long dashed line in 
Fig. \ref{fig:cross10decomp}) of the $\gamma=10$
signal, thereby enabling a better interpretation of the rest of the signal. 
 Higher multipoles
continue to 
contribute more stongly at even larger boost factors.  For example, at
 $\gamma=15$ for these calculations that terminate at the fourth multipole,
the $2^+$, $3^+$, $3^-$, $4^+$ and $4^-$ 
multipoles contribute 38\% to the total signal.
Therefore such a third measurement would be beneficial in understand
this piece, which is 
the most uncertain part of the response that would
contribute to neutrino-nucleus scattering
by supernova neutrinos.

\section{conclusions}

We have discussed the electron spectrum produced from
a lead target for two different sources of neutrinos.
Any future measurements of neutrino-nucleus cross sections
are much anticipated. However, if possible it would be desirable to
have more than one electron spectrum with which to calibrate 
theory.  We have illustrated the 
relative nuclear contribution to the
electron energy spectra for various beta beam energies and compared
this with the pion source.  We have also considered the 
contributions of the one neutron and two neutron spectrum.
One could maximize the infomation with which to compare theory
calculations by separating the spectra into parts associated with
zero, one and two neutrons, by running the beta beam at more
than one energy such as at $\gamma = 5$, $\gamma=10$ and $\gamma=15$,
or by using the electron spectra produced by both sources. 

\acknowledgements
Related work has been performed independently by \cite{newvolpe}.
G. C. M. thanks B. Flemming, R. Tayloe, and R. Hix for useful discussions.
This work was supported by DOE Grant DE-FG02-02ER41216.

\begin{table}[h]
\begin{tabular}{|c|c|c|c|} \hline
& ~~ total ~~ & ~~ one neutron ~~ & ~~ two neutrons ~~ \\
 \hline
pion source & & & \\
\hline
 $\langle \sigma \rangle \, (10^{-40} {\rm cm}^2$) &  41 & 23 & 13 \\
 $\langle {\rm E} \rangle$ (MeV) & 21 & 21 & 18  \\
Events (day$^{-1}$) & 180  & 120 & 66 \\
\hline
Beta Beam $\gamma=5$ & & &\\
\hline
 $\langle \sigma \rangle \, (10^{-40} {\rm cm}^2$) & 8.5 & 6.7 &  0.4 \\
 $ \langle E \rangle$ (MeV) & 11 & 10 & 7.1  \\
Events (day$^{-1}$) &  1.1 & 0.8 &  0.05 \\
\hline
Beta Beam $\gamma=10$ & & &\\
\hline
$\langle \sigma \rangle \, (10^{-40} {\rm cm}^2$) & 52 & 26 &  21 \\
 $ \langle E \rangle$ (MeV) &  25 &  25 & 23 \\
 Events (day$^{-1}$) & 26  & 13 & 10 \\
\hline
Beta Beam $\gamma=15$ & & &\\
\hline
$\langle \sigma \rangle \, (10^{-40} {\rm cm}^2$) & 110 & 41 & 61 \\
 $ \langle E \rangle$ (MeV) & 40 & 41 & 38  \\
 Events (day$^{-1}$) & 95 & 34 & 53\\
\hline
\end{tabular}
\caption{For various electron neutrino sources, this table
shows for $^{208}{\rm Pb}$ the flux averaged charged current cross sections,
the average energy of the electrons, and compares numbers
of events per day in the detector.  The event rates are for 100\% efficiency,
a detector with a 4m$^2$ cross sectional area, and a straight section of
beta beam track of 90m that ends 10m from the detector.  Also assumed 
are that there are $10^{13}$
decays per second in the ring, and that 90m takes up 10\% of the track.
Some electrons will
be produced in coincidence with one or two neutrons and these
numbers are also shown.\label{tab}}

\end{table}

\end{document}